\documentclass{aastex}
\usepackage{natbib,epsfig,emulateapj5,apjfonts}

\newcommand{\bm}{\mbox{\boldmath $M$}}

\newcommand{\bcd}{\mbox{\boldmath $\cal D$}}

\newcommand{\be}{\begin{eqnarray}}
\newcommand{\ee}{\end{eqnarray}}

\newcommand{\Da}{{\cal D}_a}

\newcommand{\Db}{{\cal D}_b}

\newcommand{\Lpsr}{{\cal L}}

\newcommand{\Npsr}{N_{\rm psr}}

\newcommand{\DM}{\rm DM}
\newcommand{\TMC}{T_{\rm MC}}

\begin{document}

\title{The Velocity Distribution of Isolated Radio Pulsars}
\author{Z. Arzoumanian\altaffilmark{1}}
\affil{Laboratory for High-Energy Astrophysics, NASA-GSFC,
Greenbelt, MD~20771}
\altaffiltext{1}{NRC/NAS Research Associate}
\email{zaven@milkyway.gsfc.nasa.gov}
\bigskip
\author{D. F. Chernoff}
\affil{Center for Radiophysics \& Space Research, Cornell University,
Ithaca, NY~14853}
\email{chernoff@astro.cornell.edu}
\bigskip
\author{J. M. Cordes}
\affil{Astronomy Department and NAIC, Cornell University,
Ithaca, NY~14853}
\email{cordes@astro.cornell.edu}

\bigskip
\begin{abstract}

We infer the velocity distribution of radio pulsars based on large-scale 
0.4 GHz pulsar surveys. We do so by 
modelling evolution of the locations, velocities, spins, and radio
luminosities of pulsars;
calculating pulsed flux according to a beaming model and random orientation
angles of spin and beam;  applying selection effects of pulsar surveys;
and comparing model distributions of measurable pulsar properties with
survey data using a likelihood function.  The surveys analyzed have
well-defined characteristics and cover $\sim 95$\% of the sky.  
We maximize the likelihood in a 6-dimensional space of
observables $P, \dot P, {\rm DM}, |b|, \mu, F$ (period, period derivative,
dispersion measure, Galactic latitude, proper motion, and flux density).  
The models we test are described by 12 parameters that
characterize a population's birth rate, luminosity, shutoff of radio
emission, birth locations, and birth velocities.   
We infer that the radio beam luminosity {\em (i)\/} is comparable to the
energy flux of relativistic particles in models for spin-driven
magnetospheres, signifying that radio emission losses reach nearly
100\% for the oldest pulsars; and {\em (ii)\/} scales approximately as
$\dot E^{1/2}$ which,
in magnetosphere models, is proportional to the voltage drop available
for acceleration of particles.
We find that a two-component velocity distribution with characteristic
velocities of 90 km~s$^{-1}$ and 500 km~s$^{-1}$ is greatly preferred
to any one-component distribution;  this preference is largely
immune to variations in other population parameters, such as the
luminosity or distance scale, or the assumed spin-down law.
We explore some consequences of the preferred birth velocity distribution:
{\em (i)\/} roughly 50\% of pulsars in the solar neighborhood will
escape the Galaxy, while $\sim 15$\% have velocities greater than $1000$
km~s$^{-1}$; 
{\em (ii)\/} observational bias against high velocity pulsars is relatively
unimportant for surveys that reach high Galactic $|z|$ distances, but is
severe for spatially bounded surveys; 
{\em (iii)\/} an important low-velocity population exists
that increases the fraction of neutron stars retained by globular
clusters and is consistent with the number of old objects that accrete
from the interstellar medium;
{\em (iv)\/} under standard assumptions for supernova remnant expansion
and pulsar spin-down, $\sim 10$\% of pulsars younger than 20 kyr will
appear to lie outside of their host remnants.
Finally, we comment on the ramifications of our birth velocity
distribution for binary survival
and the population of inspiraling binary neutron stars relevant to 
some GRB models and potential sources for LIGO.

\end{abstract}


\section{Introduction}

Only a small fraction of the estimated $10^9$ neutron stars (NS) in
the Galaxy are visible to Earth-bound
observers as radio pulsars. In addition to source brightness and 
survey sky coverage, many effects unique to neutron
stars determine which objects will be detected:  
(1) highly beamed emission;
(2) turning-off of radio emission as a NS spins down;
(3) migration 
from a thin 
extreme Population I $z$ distribution at birth to a combined thick disk 
and unbounded population;
(4) interstellar plasma dispersion and scattering that distort pulses,
diminishing search sensitivities to distant and rapidly rotating pulsars.
Statistical studies that aim to determine the birthrate, total number,
and spatial distribution of radio pulsars must contend with these
selection effects.  Indeed, previous studies (some recent examples are
Lyne et al.\ 1998\nocite{l+98}; Hartman et al.\ 1997\nocite{hbwv97};
Hansen \& Phinney 1997\nocite{hp97}; Lorimer, Bailes \& Harrison
1997\nocite{lbh97}) differ from one another significantly in this
respect. All analyses have been hampered by our incomplete knowledge of
intrinsic properties such as the luminosities and shapes of pulsar beams
and their variations with spin rate and magnetic field strength.

We report a new effort to derive the maximum useful information about
Galactic neutron stars from the subset observable as radio pulsars
(excluding millisecond and binary pulsars and those in globular
clusters), simultaneously solving for many of the pulsar properties
upon which the effects of survey selection depend. We use a new, highly
realistic model of NS birth, evolution, and detection in radio surveys,
and we implement a rigorous statistical analysis of the joint
distribution of observable pulsar properties drawn from real surveys.
Monte Carlo simulations of NS populations are tested for
detectability and compared, in a 6-dimensional space of observables, to
the observed pulsar sample.  We vary the model parameters to infer
most likely values and credible ranges.  To the extent that real-world
selection effects are adequately described in the model, the analysis
yields results that are independent of selection biases.

The main focus of this paper is a determination of the pulsar velocity 
distribution function at birth and its implications.
Other results, such as the luminosities of pulsar
beams and the radio pulsar birthrate, are briefly described here
(covariances do exist between these model parameters and the best-fit
velocity results), but a detailed exposition of the model, statistical
method, and insights into other areas of pulsar phenomenology will be
presented elsewhere. We summarize the essential features of our model
and method in Sec.\ 2, present our results in Sec.\ 3, and discuss some
implications of our new determination of the pulsar birth velocity 
distribution in Sec.\ 4.

\section{Model and Method}

\subsection{Summary of the Model}

We model the birth properties and rotational, kinematic, and luminosity 
evolution of a Monte Carlo population of neutron stars. The characteristics
of eight real-world surveys are then used to assess the detectability of
each simulated NS as a radio pulsar. The model-derived population is
compared to the actual survey results using a likelihood analysis.

\subsubsection{Neutron Star Birth}
\label{birth.sec}

We assume that the NS birthrate is constant in time,
uniform per unit area for a Galactic disk with 15 kpc radius, and
exponentially distributed above and below the disk midplane with
parameterized scale $z_0$.
The orientations of stellar spin and magnetic axes  
are independent and uniformly distributed over the sphere.
The birth velocity is distributed isotropically in direction
about the local standard of rest; it incorporates both an asymmetric 
``kick'' impulse from a supernova explosion and the residual
contribution from a disrupted binary, for those objects born in binaries.  
We do not separately investigate the two
phenomena here nor do we attempt to model neutron stars that remain bound in
binaries (high-field pulsars in binaries are a negligible population). 
Two model distributions of the velocity magnitude are
considered, consisting of one or two Gaussian components. The two
component model is
\begin{eqnarray}
f(v) &=& 4\pi v^2 \left[
w_1
\frac{1}{(2\pi\sigma_{v_1}^2)^{3/2}}\exp(-v^2/2\sigma^2_{v_1})
\right.
\nonumber \\
&+& \left.
(1-w_1)
\frac{1}{(2\pi\sigma_{v_2}^2)^{3/2}}\exp(-v^2/2\sigma^2_{v_2})
  \right],
\end{eqnarray}
parameterized by $\sigma_{v_1}$ and $\sigma_{v_2}$, the
dispersions of the low- and high-velocity components, respectively,
and by $w_1$, the fraction of neutron stars with parent Gaussian of
width $\sigma_{v_1}$.
Finally, the initial periods and magnetic field strengths are drawn from
log-normal distributions with parameterized means ($\langle \log P_0 \rangle$,
for $P_0$ in seconds, and $\langle \log B_0 \rangle$, for $B_0$ in Gauss),
widths ($\sigma_{\log P_0}$, $\sigma_{\log B_0}$), and fixed high and 
low cutoffs. 

\subsubsection{Evolution.}
Given initial conditions,
the evolution of a new-born pulsar to the present epoch
is deterministic and described as follows.
Most of our analysis assumes spin down according to a constant, magnetic-dipole
braking torque\footnote{
We omit from our spin-down expression the $\sin^2\alpha$ dependence of
the idealized magnetic dipole braking torque, where $\alpha$ is the
angle between the spin and magnetic field axes. In magnetosphere models
(e.g., Goldreich \& Julian 1969)\nocite{gj69}, the torque does not
vanish for aligned rotators ($\alpha = 0$).}  with ``braking index'' $n=3$.
This choice facilitates
comparison of our results with previous NS population studies, but we have
also investigated an alternate torque model, with $n =
4.5$---effects of different assumptions for $n$ are described 
in Sec.~\ref{assump.sec}.  We assume that the
direction of the magnetic field with respect to the spin axis does not
evolve with time.  Each star moves in the Galactic potential given by
Paczy\`{n}ski (1990)\nocite{pac90} for the disk and bulge, and by
Caldwell \& Ostriker (1981)\nocite{co81} for the corona. 

Each star's
luminosity is a function of period and spin-down rate, with no intrinsic
spread (i.e., pulsars are assumed to be standard candles for fixed
values of $P$ and $\dot P$),
\begin{equation}
\label{lr.eq}
L_r = \min \left\{ P^\alpha\,\dot P_{15}^\beta\,L_0, \dot E
\right\}\;{\rm erg\;s}^{-1},
\end{equation}
where $(\alpha,\,\beta,\,L_0)$ are model parameters, 
$\dot P = \dot P_{15} 10^{-15}$ s s$^{-1}$, and 
$\dot E = 4\pi^2 I \dot P/P^3$ is the spin-down energy loss rate
for moment of inertia $I = 10^{45}$ g cm$^2$.
Previous analyses have assumed intrinsic spreads in the
``pseudo-luminosity'' (flux $\times$ distance-squared) to account 
for viewing of the pulsar beam from different orientations
as well as possible differences in intrinsic luminosity.
We choose instead to explicitly model flux variations due to viewing
geometry through a beam model, described below, and a physically
meaningful luminosity (in erg s$^{-1}$).  We find that the
inevitable spread in fluxes due to viewing geometry is substantial.
Some implications of this choice are discussed in
Sec.~\ref{assump.sec}.

Slowly rotating neutron stars do not emit in the radio because
the electric field near the polar cap is too small (e.g.,
Chen \& Ruderman 1993)\nocite{cr93}. The cutoff may not be abrupt;
we implement a ``death band'' as
the probability that objects with given ${\dot P}/P^3$ are radio
quiet. The probability is
\[ 
1/2\left\{ \tanh \left[ 
(\dot P_{15}/P^3 - 10^{\overline{\rm DL}})/\sigma_{\rm DL} 
			 \right] + 1 \right\} , 
\]
for $P$ in seconds, so that
the position and lateral extent of the death band are
parameterized by $\overline{\rm DL}$ and $\sigma_{\rm DL}$.

\subsubsection{Beaming and Pulse profiles.}
\label{beam.sec}

For each simulated neutron star, a pulse profile is generated using a
phenomenological radio-beam model.  Our beam model was fixed by fitting
0.4 GHz pulse profiles to ten known pulsars with viewing geometries that
are well-determined from polarization measurements (e.g., Lyne \&
Manchester 1988; Rankin 1993)\nocite{lm88,ran93}, and which span much of
the $P$-$\dot P$ region of interest: PSRs B0301+19, B0329+54,
B0450$-$18, B0525+21, B1541+09, B1642$-$03, B1821+05, B1933+16,
B2002+31, B2111+46. 
Satisfactory fits were achieved with the following beam description,
which (together with our view that pulsars are standard candles for
given $P$, $\dot P$) is consistent with preliminary results of a more
ambitious analysis of pulse waveforms and luminosities to be presented
elsewhere. 

\begin{figure}
\epsscale{0.8}
\plotone{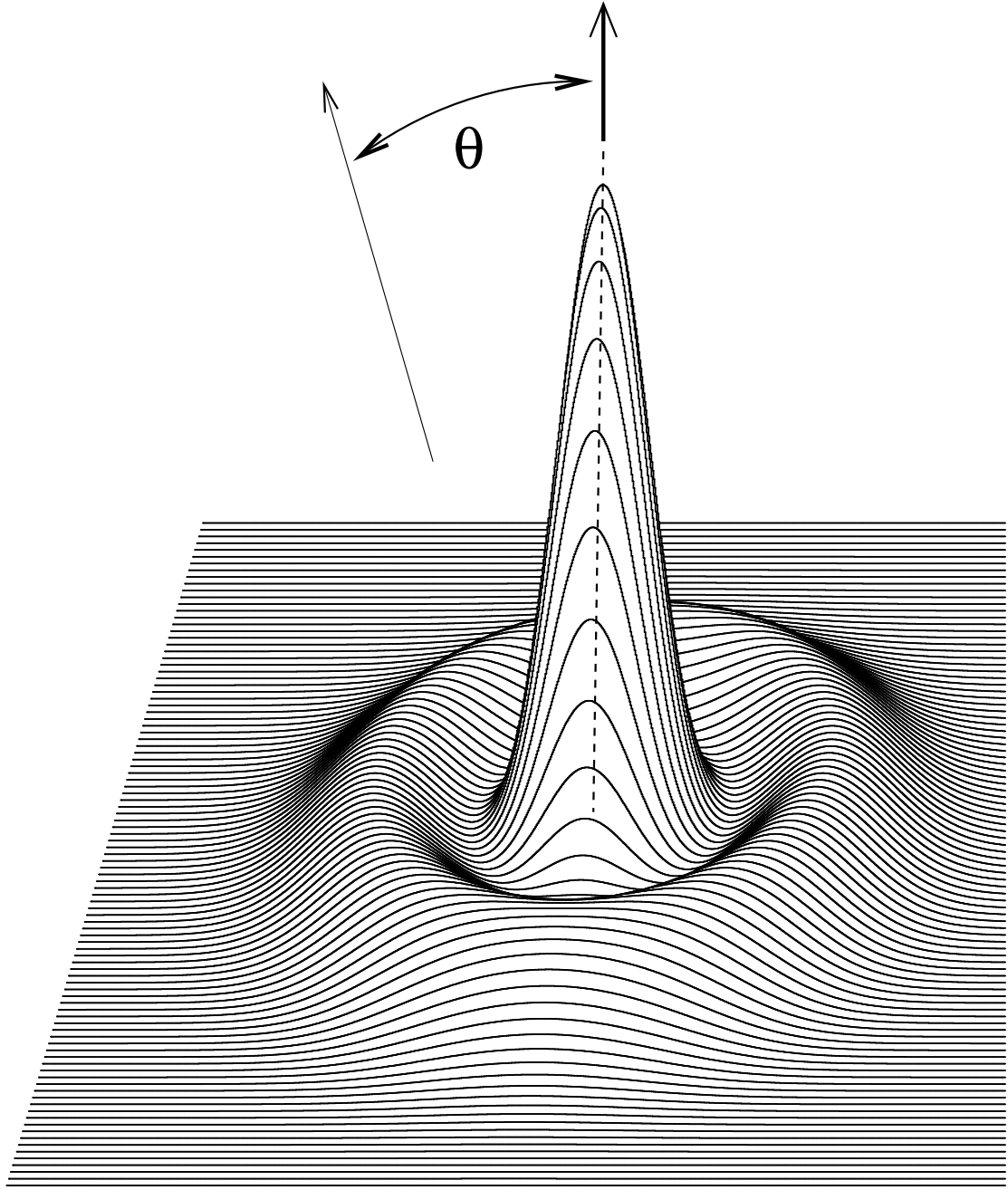}
\caption{\label{beam.fig} Illustration of 0.4 GHz radio flux distributed in
the core and conal components of our beam model, for $P = 1$ s.}
\end{figure}

In our model, pulsed radio flux from a NS depends on contributions from one
``core'' and one ``conal'' beam. Each is Gaussian in form with opening
angle $\rho$ (half-width at $1/e$) that scales with period as
\begin{eqnarray}
\rho_{\rm core} & = & 1.5^\circ P^{-0.5} \\
\rho_{\rm cone} & = & 6.0^\circ P^{-0.5}.
\end{eqnarray}
The beams subtend solid angles $\Omega = f \pi \rho^2/\ln 2$, 
where the correction factor $f_{\rm cone} = 0.4$ accounts for 
the hollowness of the cone beam, and $f_{\rm core} = 1$. 
The total radio luminosity $L_r$ is apportioned 
such that the ratio of core to cone peak flux is 
\begin{equation}
F_{\rm core}/F_{\rm cone} = {\textstyle\frac{20}{3}} P^{-1}.
\end{equation}
The core increases in strength with decreasing period and
decreasing observing frequency $\nu$; the latter dependence is not
relevant to the present analysis of 0.4 GHz surveys.
An example of the
distribution of core and conal flux is shown in Fig.~\ref{beam.fig}.

\begin{figure}
\epsscale{0.8}
\plotone{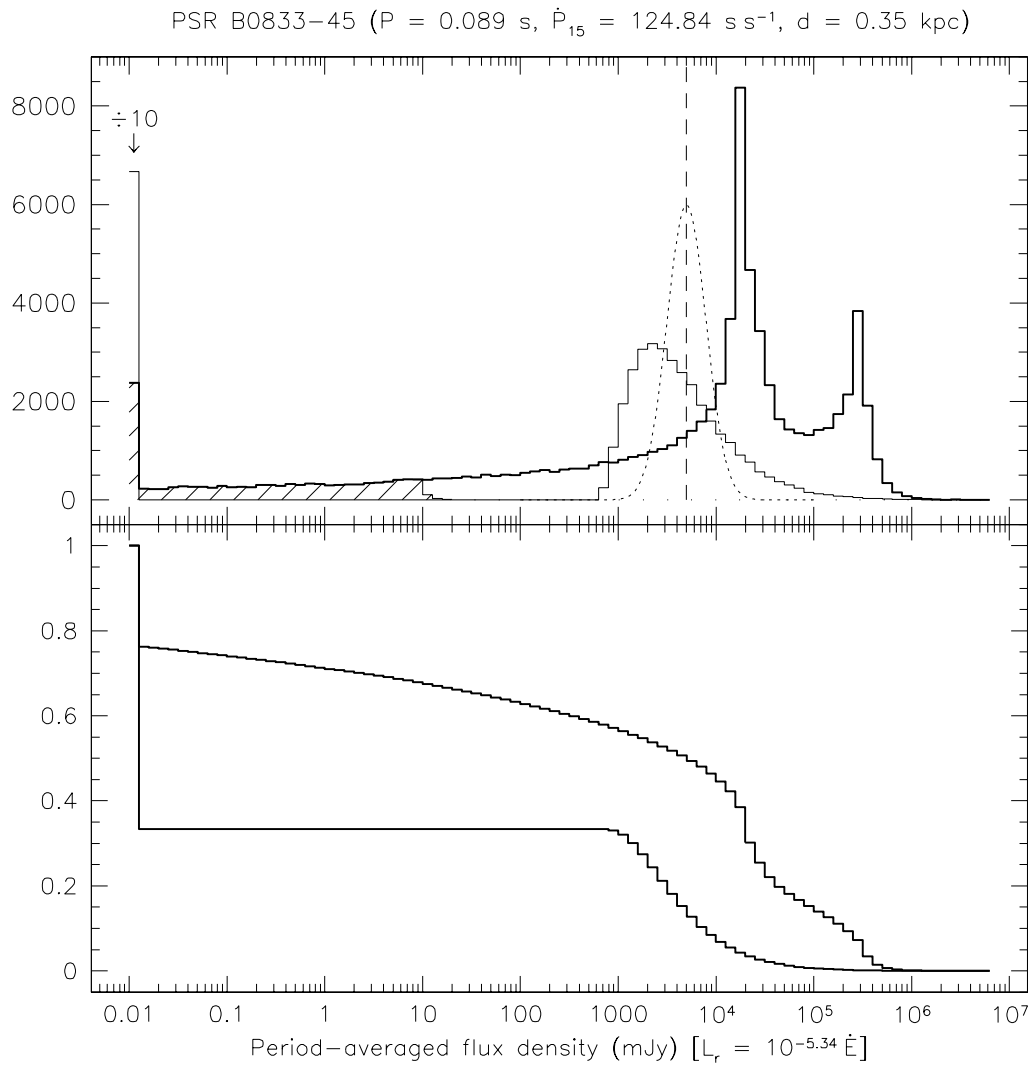}
\caption{\label{pflux.fig} 
The differential ({\em top}) and cumulative ({\em bottom}) distributions
of period-averaged flux arising from: ({\em thick line}) our beam and best-fit 
luminosity
models, eq.~\protect\ref{lrresult.eq}, given the period, spin-down rate, and
distance of the Vela pulsar, for $10^5$ randomly-drawn viewing geometries;
({\em thin line}) the spread in ``pseudo-luminosity'' used
by Hartman et al.\ (1997), their Eqs.~(2) and (3). 
The dashed line indicates the pulsar's cataloged flux; the dotted
Gaussian curve depicts the likelihood function we use to represent
measurement error. See text, Sec.~\ref{assump.sec}.
For our beam description, the hatched histogram
represents viewing configurations that are undetectable by the
Parkes 0.4 GHz survey (Manchester et al.\ 1996);
the cumulative distribution 
indicates that a similar all-sky survey with a 10 mJy flux
limit will detect two-thirds of the Vela-like pulsars
at the 350 pc assumed distance.}
\end{figure}

Pulse waveforms are constructed for the viewing geometry of each Monte
Carlo pulsar, with two oppositely-directed radio beams centered on the
star's magnetic axis.  The waveform flux at pulse phase $\phi$
corresponding to polar angle $\theta(\phi)$ to the line of sight is
\begin{equation}
F(\phi) = 
F_{\rm core}\exp\left\{-\theta^2/\rho_{\rm core}^2\right\} + 
F_{\rm cone}\exp\left\{-(\theta-\bar{\theta})^2/w_e^2\right\},
\end{equation}
where $\bar{\theta}$ and $w_e$ 
are the centroid and effective width of the Gaussian conal beam. 
Figure~\ref{pflux.fig} shows typical differential and cumulative
distributions of period-averaged flux density that result from our beam model
and radio luminosity inferred from the likelihood analysis
(Sec.~\ref{results.sec}). 
The highest fluxes correspond to nearly aligned viewing geometries, 
while the tail at low flux corresponds to lines of sight grazing the
skirts of the radio beam.  The two peaks in the differential
distribution arise from the core (narrower beam, higher average flux)
and cone (wider beam, lower average flux) components. 

\subsubsection{Detection.}

The effects of interstellar dispersion and scattering are applied to the
pulse profile of each potentially detectable NS, where the degree of
waveform ``smearing'' depends on the applicable survey instrumentation.
The DM and SM values used are based on the Taylor \& Cordes (1993)
model, dithered to simulate errors in the DM-distance relationship at
the 50\% level.  Pulse profiles are Fourier transformed and tested for
detection by each survey with sky coverage in the direction of the
simulated pulsar. The minimum detectable flux is computed for the
antenna gain, receiver and sky temperature, bandwidth, dwell time,
sample interval, and detection threshold (including harmonic summing) 
applicable to each survey. 
Detection probabilities were modified by the ratio of successfully
searched area to the area contained within the survey boundaries on
the sky. We do not
model specific telescope pointings, but assume uniform antenna gain
across the region surveyed by each telescope. In reality, antenna gain
drops off away from the pointing direction, and survey pointings are
generically arranged in a close grid with adjacent pointings separated
by the half-power width of one, roughly Gaussian, telescope ``beam.''
The assumption of constant gain results in an enhanced detection rate
and, hence, a lower inferred birthrate, which we later correct by the ratio
of volumes for beams with constant and Gaussian cross-sections,
$f_{\rm gain} = 1.4$.

\smallskip
In summary, we draw neutron stars independently from 
a parameterized parent population, evolve
them according to the simulation model and test for detectability as
pulsars in the given surveys.  This ``forward evolution'' from birth
to detection yields distributions of simulated pulsar properties 
that may be compared to those of the observed sample.

\subsection{Surveys and Data: Selection Criteria}

The surveys modeled, all at or near 0.4 GHz, are those of Camilo, Nice
\& Taylor (1996)\nocite{cnt96}, Dewey et al.\ (1985)\nocite{dtws85},
Foster et al.\ (1995)\nocite{fcwa95}, Manchester et al.\
(1978)\nocite{mlt+78}, Manchester et al.\ (1996)\nocite{mld+96}, Nice,
Fruchter \& Taylor (1995)\nocite{nft95}, Stokes et al.\
(1985)\nocite{stwd85}, and Thorsett et al.\
(1993)\nocite{tdk+93}. Together, they cover some 95\% of the
sky. The data are culled from the May 1995 electronic version of
the Taylor et al.\ (1993; 1995) catalog of 721 pulsars, supplemented
with 66 $\dot P$ measurements from D'Amico et al.\ (1998), including
the 8.5 sec pulsar of Young et al.\ (1999). 
Of the full set of
known pulsars we keep only those that were detected in at least one of 
these surveys; for example, we exclude objects found in deep
searches that targeted supernova remnants or X-ray sources, and pulsars 
detected only in high-frequency (1.4 GHz or higher) searches. 
In addition, to avoid contamination from objects
with complicated evolutionary histories, we exclude from our analysis:
binary pulsars, pulsars in globular clusters, extragalactic (LMC and
SMC) pulsars, and all pulsars with spin-down rates less than
$10^{-16}$ s s$^{-1}$ because of potential confusion with
spun-up objects.  The resulting dataset contains 
$\Npsr = 435$ pulsars, 79 of
which have measured proper motions or useful upper bounds. An additional
54 cataloged pulsars meet the criteria above but do not have measured
spin-down rates. We account for these objects in our derived birthrate
through a correction factor $f_{{\rm no}\,\dot P} = (435+54)/435$,
but otherwise neglect them in the likelihood analysis.

\subsection{Likelihood Function}
\label{like.sec}


The distributions of measured quantities that characterize pulsars
arise from variations intrinsic to the NS population, bias in survey
selection, and measurement errors. We account for measurement errors
as reported in the literature, and
model intrinsic variations through Monte Carlo (MC) realizations
subject to the same detection criteria used to obtain the 
observed pulsar sample. Measurable quantities (``observables'') are
of two types: 
{\em (a)} $\Da = \left\{ P,\dot P,|b|,\DM \right\}$, 
    which are known with sufficient precision that their distributions
    are determined by the biased sampling of the population; and
{\em (b)} $\Db = \left\{\mu, F\right\}$, 
    proper motion and flux density, which carry significant measurement
    uncertainty described by observer-reported or assumed likelihood
    functions ${\cal L}^{\rm obs}_{\left\{\mu,F\right\}} \equiv {\cal L}^{\rm
    obs}_\mu {\cal L}^{\rm obs}_F$. We assume a Gaussian form for
    ${\cal L}^{\rm obs}_\mu$, and log-normal distributions above a
    detection threshold for ${\cal L}^{\rm obs}_F$.

We do not know a priori the form of the likelihood function for $\Da$,
and therefore construct it numerically. We first partition the
4-dimensional space for $\Da$ using the real sample of pulsars to define
bins in a heirarchical fashion---each bin corresponds to a small volume
element centered on specific values of $P,\dot P, |b|,$ and DM.
We then create MC pulsars according to a specific 
population/evolution/luminosity model, and
find the total weight $W_i$ of detected objects with observables that fall
within the boundaries of the $i$-th bin. For birthrate $\dot N$, the
expected number of such objects is $\langle n_i \rangle = \dot N
(\TMC/W)W_i$. Here, $W$ is the total weight of MC neutron
stars born uniformly distributed over time interval $\TMC$.
The likelihood for the $\Da$ dataset is then formed using products of
Poisson probabilities for 
$0$ or $1$ elements in each bin, 
\begin{equation}
\label{la.eq}
{\cal L}_a 
	     = e^{- \langle n \rangle} \prod_i \langle n_i \rangle,
\end{equation}
where the product is restricted
to occupied bins, and the total expected number of pulsars is 
\begin{equation}
\langle n \rangle = \sum_i \langle n_i \rangle.
\end{equation}
For precisely measured quantities, the numerical evaluation of
${\cal L}_a$ is exact in the limit of zero bin size and small occupancy.
In practice, we use bins that
enclose more than one data pulsar, and we generalize the likelihood
function accordingly. 

Information from $\Db$ observables must also be included.  The weight
of Monte Carlo points in the $i$-th bin is $W_i = \sum_k w_k$ (where $k$
counts individual MC points); the weight of points consistent with the
observer-reported uncertainties is
$\sum_k w_k {\cal L}^{\rm obs}_{\left\{\mu,F\right\}_k}$ 
(sum over joint likelihood of proper motion and flux  for data in the
$i$-th bin),
which may be rewritten as $W_i \left( \sum_k w_k {\cal
L}^{\rm obs}_{\left\{\mu,F\right\}_k} / \sum_k w_k \right) = W_i
\langle \langle {\cal L}^{\rm obs}_i \rangle \rangle$. The expected
number of pulsars in the $i$-th bin consistent with the flux and proper
motion observations is ${\dot N} (\TMC/W) W_i \langle
\langle {\cal L}^{\rm obs}_i \rangle \rangle$.  The likelihood of the
whole dataset of precisely and imprecisely measured quantities is then 
${\cal L} = e^{- \langle n \rangle} \prod_i
\langle n_i \rangle \langle \langle {\cal L}^{\rm obs}_i \rangle \rangle$.

Maximizing $\Lambda = \ln {\cal L}$ with respect to ${\dot N}$ yields
the birthrate
\begin{equation}
\label{br.eq}
{\dot N_{\rm ML}} = \left( N_d \over W_d \right) \left( W \over \TMC
\right),
\end{equation}
where $N_d = 435$ is the number of actual pulsars, $W_d$ is the weight of
{\em detected\/} Monte Carlo points, $W$ is the total weight of
Monte Carlo points and $\TMC$ is the timescale of the simulation.
In this paper we analyze the model parameters using the likelihood
evaluated at the most likely birthrate $\dot N_{\rm ML}$:
\begin{equation}
\label{like.eq}
\Lambda|_{\dot N_{\rm ML}}  = 
\sum_i \ln \left( W_i \over W_d \right) +
\sum_i \ln \langle \langle {\cal L}^{\rm obs}_i \rangle \rangle,
\end{equation}
where a constant term has been omitted.

Given the likelihood function (eq.~\ref{like.eq}), we employ the
techniques of Bayesian inference to derive plausible ranges
for model parameters, and to compare models with different
numbers of parameters.  (For a review of the essentials of Bayesian
inference, see Gregory \& Loredo 1992\nocite{gl92}.)
Bayes's theorem states that the posterior probability distribution of a
model $\bm$ given a dataset $\bcd$ is proportional to the probability of
the data given the model, $P(\bcd | \bm, I)$---i.e., the likelihood
${\cal L}$---times the prior probability of the model $P(\bm | I)$,
where $I$ represents assumptions within the model.  We assume flat
priors over finite ranges of the model parameters, so that the peak of
the posterior is just the maximum of ${\cal L}$. 
We derive an uncertainty estimate (or ``credible region,'' in the
parlance of Gregory \& Loredo) for each model parameter
from the marginal probability distribution for that parameter, i.e.,
from the likelihood function integrated over all other parameters.
Similarly, a global likelihood {\em for each model\/} may be derived by
integrating over all model parameters---such global likelihoods offer a
rigorous means of comparison between models that have different
degrees of complexity as reflected in the number of model parameters.
Ratios of global likelihoods between models are known as ``odds
ratios.'' 

\subsection{Implementation}

For the $\Da$ observables, we defined a 4-dimensional space 
with 4 partitions in each dimension. With this choice, the 435 known
pulsars were distributed into 164 distinct bins, filling 64\% of the
observable phase space. 
We found that the addition of a fifth dimension, Galactic
longitude $l$, to the space of observables, or the use of two
orthogonal components of proper motion ($\mu_l$ and $\mu_b$, say)
rather than the magnitude, yielded too few MC points in some bins. 

We tested our method by generating simulated pulsar data ``catalogs''
according to a particular luminosity and velocity model and attempting
to recover the model parameters. The mock datasets were constructed to
resemble our true dataset, both in the number of pulsars and in the
number and quality of proper-motion measurements.  We found that the
resulting maximum likelihood parameter values differed from their 
expected values by no more than 20\%, with the 99\% ``credible region'' 
of the marginalized posterior probability distribution for each parameter 
(see Sec.~\ref{like.sec}) implying uncertainties of 10\% for the luminosity
parameters $\beta$ and $\log L_0$, 15\% for $\alpha$, and 30\%
for each of the velocity parameters $(\sigma_{v_1},\sigma_{v_2},w_1)$. 
The model that described the mock
dataset was reliably contained, in a half-dozen runs of the simulation,
within the volume of the likelihood function that contained at most 90\%
of the probability.  Further details on mock data recovery tests will be
presented elsewhere.  

For the likelihood analysis, we generated a Monte Carlo dataset of
nearly $10^{10}$ pulsars for $\TMC = 1$ Gyr, drawing the points from a
large range of parameter values; the large value of $\TMC$ was chosen to
readily accomodate the oldest (characteristic age $\sim 300$ Myr) cataloged
pulsars and a range of braking models.  We tested each point for
detection in any one of the eight surveys and used the resultant bin
occupancies to derive the likelihood function, $\Lpsr$, as defined
above.
We tested the numerical stability of $\Lpsr$ for the best models by
increasing the number of MC points.

Computations were carried out on the 136-node IBM SP2
supercomputer at the Cornell Theory Center; evaluation of a
single set of model parameters typically required 2.5 minutes on 48
nodes running in parallel. Computational limitations precluded a
full 12-dimensional grid search. Instead, we evaluated models on 
12-dimensional ``crosses,'' iteratively repositioning the cross
in search of a global maximum. We also searched over 3- and 5-dimensional 
sub-grids of the full parameter space, exploring in particular 
the velocity $(\sigma_{v_1}, \sigma_{v_2}, w_1)$ and luminosity
$(\alpha, \beta, L_0)$ parameters. We have found a well-defined
global maximum in the posterior probability function for 
models given the data.

\section{Results}
\label{results.sec}

\subsection{Birth Properties}
Table 1 displays the model parameter values that maximize the posterior
probability for models given the data (with flat priors) over the indicated
ranges.  The Galactic birthrate of NSs visible as radio pulsars resulting from
eq.~\ref{br.eq} is 
\begin{equation} 
\label{ndot.eq}
\dot N = f_{\rm gain} \times f_{{\rm no}\,\dot P} \times \dot N_{\rm ML} =
(760\,{\rm yr})^{-1}.
\end{equation}
The first two factors in eq.~\ref{ndot.eq} represent the only parts of
the analysis that are not explicitly modeled by Monte Carlo, i.e., the
corrections for cataloged pulsars with
unmeasured spin-down rates and for variations in antenna gain as described
above. These should not be confused with the more typical beaming
corrections. Because our model incorporates 
a specific description of the radio beam, no {\em post facto\/} beaming
correction is needed (or allowed), as it is in previous studies.
Nonetheless, the value of $\dot N$ under-represents the true Galactic
radio pulsar birthrate.
As shown in Table 1, we impose
cutoffs on the birth magnetic field ($B_0$) distribution corresponding
to the highest and lowest field strengths in our sample of cataloged
pulsars; any unseen pulsars lying outside of this range have not been 
accounted for by our $\dot N$. The value of $\dot N$ will also depend
on details of the radial distribution of NS birthsites in the Galactic disk,
which we have fixed in our analysis (Sec.~\ref{birth.sec}), on our
description of the pulsar beam (e.g., the assumption of circularity;
Sec.~\ref{beam.sec}), and on the assumed spin-down law. We compare
our birthrate with previous estimates in Sec.~\ref{isol.sec}.

The distribution of spin periods at birth in our model is poorly
constrained by the data. It is consistent with birth at rapid rotation
rates (as dictated by the presence of the Crab and Vela pulsars in our
dataset), but the assumed uni-modal (log-normal) form for the
distribution cannot test the hypothesis of ``injection'' of pulsars born
with long rotation periods (Narayan \& Ostriker 1990\nocite{no90}). The
birth field-strength distribution, by
contrast, is very well constrained, with the mean of the log-normal
distribution roughly consistent with the (present-day) mean field-strength for
our dataset ($10^{12.1}$ G) as expected, given our assumption of no
torque decay.  Cut-offs in the $P_0$ and $B_0$ distributions at low and
high values were fixed for all models, as shown in Table 1. Finally,
the preferred birth scale-height in our model is $z_0 \sim 160$ pc,
consistent with the results of Cordes \& Chernoff (1998)\nocite{cc98}.

\subsection{Radio Luminosity}
The luminosity law with maximum posterior probability is 
\begin{equation}
\label{lrresult.eq}
L_r = P^{-1.3} \dot P_{15}^{0.4} 10^{29.3}\;{\rm erg\;s}^{-1},
\end{equation}
for emissions in the
radio band ($\nu > 50$ MHz) with power-law flux spectra. Comparison of
our result with earlier statistical analyses is made difficult by the
fundamental differences between our model and previous descriptions of
pulsar luminosities.  Note, however, that for luminosity proportional to
magnetic field strength, the exponents would take on the values
$(\alpha,\beta) = (0.5,0.5)$, while scaling with the spin-down energy
loss rate would imply $(\alpha,\beta) = (-3,1)$. We find instead that
the luminosity scaling with spin parameters is similar to that 
for the vacuum potential drop across the magnetic polar cap of a neutron
star (e.g., Goldreich \& Julian 1969)\nocite{gj69}, $L_r \propto \dot
E^{1/2}$.

$L_r$ is comparable in value to 
particle energy loss rates $\dot E_p$ calculated in ``polar cap'' models
for the magnetosphere. For example, the energy flux in particles
derived by Ruderman \& Sutherland (1975; their Eq.~29) may be
written as $\dot E_p = P^{-1.71} \dot P_{15}^{0.43} 10^{30.0}$
erg s$^{-1}$, while Arons \& Scharlemann (1979)\nocite{rs75,as79}
construct a ``diode'' model which implies 
$\dot E_p = P^{-1.45}\dot P_{15}^{0.4} 10^{31.2}$ erg s$^{-1}$. 
Comparison of $L_r$ with $\dot E_p$ suggests that the efficiency of radio
emission alone (i.e., not including X- and $\gamma$-ray emission)
approaches unity as pulsars age.  A more model-independent statement can
also be made: it is conceivable that the death band is related to an
upper bound in the efficiency with which spin energy is converted to
radio emission.  If we write $L_r = \epsilon_r \dot E$ $(\epsilon_r \le
1)$, our luminosity law implies an efficiency of at least $\epsilon_r =
2$\% for pulsars near
the point in the $P$-$\dot P$ diagram where they appear to shut off
($P\sim 1$ s, $\dot P_{15} \sim 0.1$); the largest implied efficiency
among the pulsars in our dataset is $\epsilon_r = 30$\% for J2144$-$3933, 
the most slowly-rotating radio pulsar known ($P = 8.5$ s; Young et al. 1999).
The requirement that $L_r \leq \dot E$
implies a lower bound $\dot P_{15}\ge 0.10\,\epsilon_{0.02}^{-1.7}\,
P^{2.8}$, a scaling with period not unlike the death band we have
assumed in our model. 

We have chosen a death-band slope of 3 ($\log\dot P_{15} = 3\log P +
\overline{\rm
DL}$ describes the center of the death-band) in order to constrain the
efficiency of radio luminosity. Theoretical expectations for the slope,
however, generally lie between 2 and 3 (e.g., Zhang, Harding \& Muslimov 
2000\nocite{zhm00}).  To investigate possible dependence of the form of $L_r$
on the assumed death-band slope, we have evaluated luminosity models
(maximizing the likelihood over a grid in $(\alpha,\beta,L_0)$ with all other
parameters fixed) for a handful of death-band slopes in this range---we
find that the maximum-likelihood values of the luminosity parameters 
remain essentially unchanged. We will parameterize the death-band slope
in future work. For the assumed $\dot P/P^3$ line, the maximum likelihood 
model implies $\overline{\rm DL} = 0.5$, and $\sigma_{\rm DL} = 1.4$.

Finally, we note that the high-energy ($h\nu \gtrsim 1$ eV) luminosities
of seven pulsars detected with the EGRET $\gamma$-ray telescope also
seem to scale as $\dot E^{1/2}$ (Thompson et al.\ 1999 and references
therein)\nocite{djt+99}.


\subsection{Birth Velocities}
\label{vel.sec}
\begin{figure}
\epsscale{0.8}
\plotone{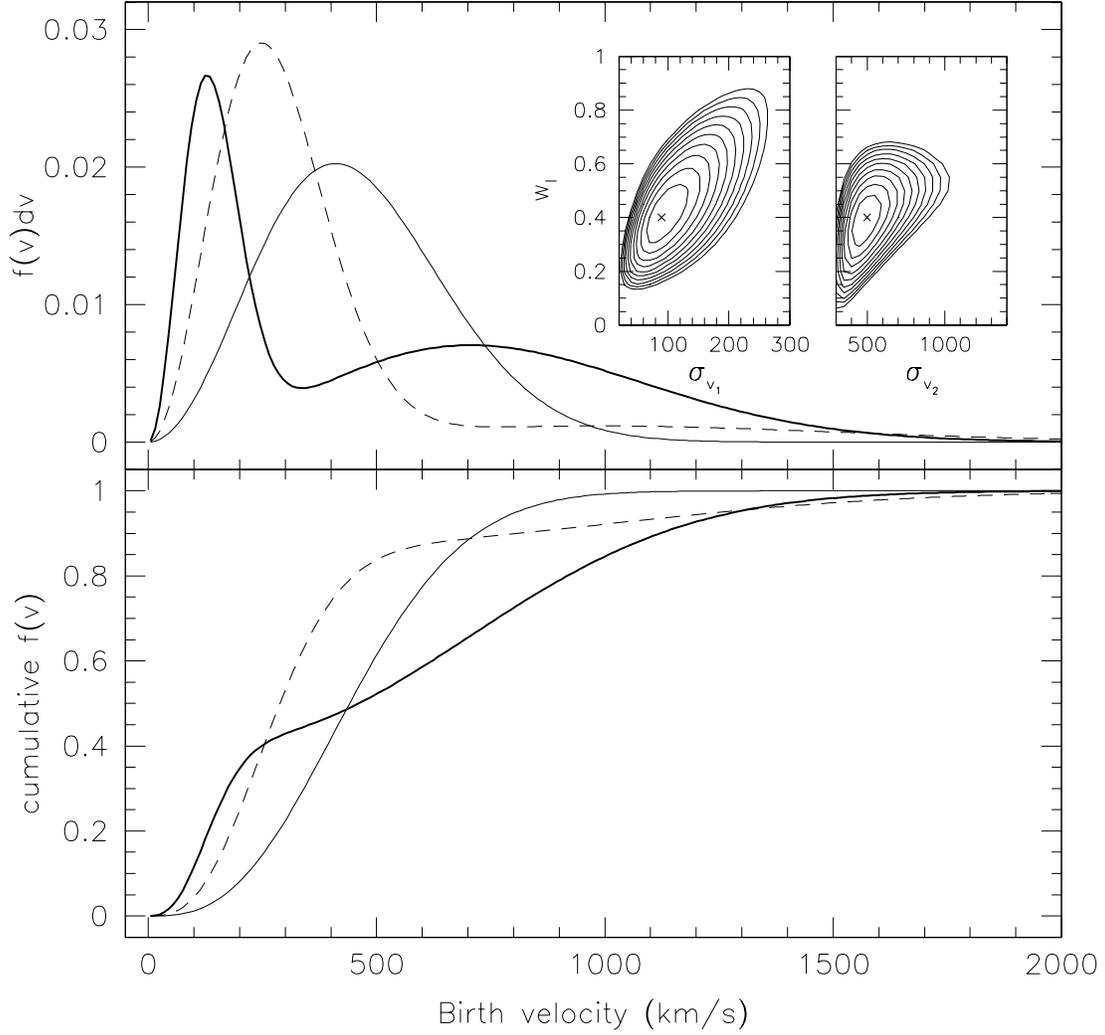}
\caption{\label{vdist.fig} Differential and cumulative distributions of 
birth velocities of NSs. The heavy line represents the
maximum-likelihood two-component distribution, with three-dimensional 
dispersions of the
Gaussian components $\sigma_{v_1} = 90$ km~s$^{-1}$ and $\sigma_{v_2} =
500$ km~s$^{-1}$, with $w_1 = 40$\% of stars drawn from 
the low-velocity component. The thin line represents the best
one-component Gaussian model, with dispersion $\sigma_v = 290$
km~s$^{-1}$. These results hold for magnetic dipole braking and no
field decay. The dashed line represents the best-fit velocity
distribution determined by Cordes \& Chernoff (1998; $\sigma_{v_1} =
175$ km~s$^{-1}$, $\sigma_{v_2} = 700$ km~s$^{-1}$, and $w_1 = 86$\%) 
from a sample of young pulsars, without correcting for selection effects 
in pulsar surveys. {\em Inset:} Likelihood contours for the velocity
parameters (all others held fixed at their maximum-likelihood values). The
contour interval is $\Delta \ln \Lpsr = -2$ throughout.}
\end{figure}

The maximum likelihood one- and two-component birth velocity distributions are
displayed in Fig.~\ref{vdist.fig}, together with contours of log-likelihood
evaluated in the vicinity of the peak.  For the two-component model,
$\sim 50$\%
of pulsars have 3-dimensional velocities greater than 500 km~s$^{-1}$, and
about 15\% have $v > 1000$ km~s$^{-1}$; also, 10\% have velocities less
than 100 km~s$^{-1}$.
For the
single-component model, only a small fraction of stars have 3-D velocities
greater than 1000 km~s$^{-1}$ or less than 100 km~s$^{-1}$. As is
evident in Fig.~\ref{vdist.fig}, the two-component distribution cleanly
differentiates fast from slow pulsars, with a deficit of objects at 
intermediate velocities (300--400 km~s$^{-1}$).
We compare the one- and two-component model
likelihoods using a Bayesian odds ratio, marginalizing the posterior
probability over the one and three parameters, respectively, to determine
global likelihoods for the models.  We find that the two-component model is
favored $\sim 10^4$ to one.  

Our best one-component velocity model is similar to that 
of Lyne \& Lorimer (1994)\nocite{ll94}, but the true distribution is
apparently poorly described by a single component. Instead,
the two-component distribution preferred by our analysis produces a
significantly larger fraction of high-velocity pulsars (and a larger mean
velocity, $\sim 540$ km~s$^{-1}$) than has been required by any previous study
of the radio pulsar population. We believe this is a consequence of (1) the
unbiased nature of our analysis, in contrast to previous efforts that have not
accounted fully for observational selection effects, and (2) our inclusion of
deep, high-latitude surveys such as the Parkes 0.4 GHz survey of the southern
sky (Manchester et al.\ 1996); see Sec.~\ref{comp.sec} for a detailed
discussion of these effects.

We have explored models that include a third,
very low velocity ($\leq 50$ km~s$^{-1}$) component. 
We find that such models produce only marginally improved likelihoods,
and acceptable models permit at most 5\% of NSs in such a low-velocity
sub-population. 
In terms of a Bayesian odds ratio, the small improvements in the
overall likelihood are insufficient to support the added complexity of 
models that include a third velocity component.
This result is similar to the finding of Cordes \& Chernoff
(1998)\nocite{cc98} but is more significant 
because our detection model accounts for selection against pulsars deep 
in the dispersing and scattering medium of the Galactic disk, to which 
low-velocity pulsars would be confined.

\smallskip
As shown in Table 1, we have also evaluated models that assume $n =
4.5$ and no torque decay. The maximum likelihood of such models is
smaller than the maximum likelihood for $n = 3$ by three orders of
magnitude for the same number of parameters, implying that canonical
dipole spin-down is preferred over this one alternative braking model.
We note, however, that the luminosity parameters $(\alpha, \beta,
L_0)$ do not change significantly, nor does the mean birth magnetic
field strength $\langle \log B_0 \left[{\rm G}\right]\rangle$. The
impact of braking-law assumptions is treated in more detail in the
following section.

\subsection{Effects of Model Assumptions}
\label{assump.sec}

The maximum likelihood (ML) parameter values shown in Table 1 are valid
in the context of assumptions made in our model---i.e., the quoted
uncertainties are purely statistical,
reflecting the 68\% credible region defined by the likelihood function.
Two fundamental assumptions may be particularly relevant to the our
birth velocity results: spin-down through
magnetic dipole braking, with no field decay, and the standard-candle
hypothesis of pulsar luminosities. We discuss the potential effects
of these assumptions below.

\subsubsection{Dipole ($n = 3$) braking.}
Because our likelihood function 
matches multiple simulated and observed pulsar properties simultaneously, at
least two sources of velocity information are available: (1) the
distribution of present-day Galactic scale heights of radio pulsars together
with some estimate of their ages, and (2) proper motion measurements.  
We find that the spatial distribution alone significantly constrains
birth velocities: if we ignore proper motion information, 
the peak of the posterior probability occurs
at $\sigma_{v_1} = 80$ km~s$^{-1}$, $\sigma_{v_2} = 350$ km~s$^{-1}$,
and $w_1 = 0.40$, consistent with the outcome based
on the full likelihood, but with a smaller average velocity.
Measured proper motions contribute preferentially to the high-velocity
tail of the distribution because some middle-aged or old pulsars at
low Galactic $|z|$ can masquerade as low-velocity objects until a proper
motion measurement reveals that they are moving rapidly---the
Guitar Nebula pulsar is a relevant example (Cordes, Romani \& Lundgren
1993)\nocite{crl93}. 

Because spin-down governs the timescale on which neutron stars leaving the 
Galactic disk remain active as radio pulsars, the choice of braking
index must affect the ML velocity model.  We may qualitatively
assess the effects of different spin-down laws and field decay as follows.
Braking index measurements for the Crab and two other young pulsars yield
values $n < 3$; if this holds also for older objects, spin-down proceeds more
slowly than we have modeled and, for a given pulsar, a lower characteristic 
birth velocity would be
needed to reach the same Galactic $|z|$-height. Field decay, however, works in
the opposite sense: if pulsar luminosities decay with decreasing field
strength, as seems likely, field decay shortens the time in which an
active pulsar must travel from its birthplace to its present-day
vertical distance from the Galactic disk, and larger velocities than
those we have derived would be necessary. The formal ML result in the
analysis of Cordes \& Chernoff (1998) for braking index and field-decay
timescale $\tau_B$ suggests $n \simeq 2.5$ and $\tau_B \simeq 6$ Myr, or
(with similar likelihood) $n \simeq 4$--5 and no field decay.  The values
of $n$ and $\tau_B$ in the first case would tend to alter our result in
competing directions, while the rapid spin-down implied in the second
case further enhances the high-velocity fraction in the derived velocity
distribution, as demonstrated by our best model with $n = 4.5$ (the mean
velocity increases from 540 km~s$^{-1}$ to 660 km~s$^{-1}$).
Furthermore, the proper motion measurements that boost the
high-velocity component are not directly affected by details
of the chosen spin-down law.  We believe, therefore, that our
primary conclusion, namely that a large fraction of Galactic NS are born
with high velocity, is robust. 

\subsubsection{Pulsars as standard candles.}
A unique aspect of our simulation is the description of observed
pulsar fluxes as an explicitly modeled consequence of viewing
orientation across a well-defined radio beam with intrinsic
luminosity that depends only on period and spin-down rate, with no
intrinsic spread. Past practice has been to allow for a range of
observed fluxes (for fixed distance) from an ad-hoc distribution
intended to mimic the combined effects of beaming orientation and
any intrinsic spread in luminosity. Our explicit beaming model is a
physically well-motivated improvement on past practice, and the
standard-candle (for given spin parameters) hypothesis is the
simplest allowed by the data. We investigate the possible
consequences of such a fundamental change in the treatment of pulsar
luminosities as follows.

In the top panel of Fig.~\ref{pflux.fig}, we show the distribution
of observed fluxes, for a Vela-like young pulsar, that result from
our beam and best-fit luminosity model ({\em thick line}) and from
the smeared pseudo-luminosity distribution of Pr\'{o}szy\'{n}ski \&
Przybycie\'{n} (1984)\nocite{pp84} as modified by Hartman et al.\
(1997; see their Eqs.\ 2 and 3)\nocite{hbwv97} ({\em thin solid
line}), where a beaming fraction of 3 has been assumed for the
latter, i.e., two-thirds of the simulated pulsars are assumed to be
beaming away from the line of sight. Several points of comparison
between the two flux distributions are noteworthy. First, the
explicit beam model produces observed fluxes spanning a wider range
than does the Hartman et al.\ smearing function, a somewhat
surprising result that depends, of course, on the chosen shape of
the beam (the outer ``conal'' component falls off as a Gaussian in our
model). Second, the proportion of undetectable stars differs between
the two models by less than a factor of two, a reflection of their
effective beaming fractions.  The difference is primarily due to the
low-flux tail of the beamed distribution, detectable between 10 and
200 mJy---if the beam skirts were to fall off more rapidly than a
Gaussian, the detection efficiency in our simulation would decrease,
resulting in a higher birthrate through Eq.~\ref{br.eq} (see
Sec.~\ref{isol.sec}). Third, when convolved by the Gaussian
representing measurement error ({\em dotted curve}), the shape of
the beamed distribution begins to resemble that of the distribution
adopted to represent an intrinsic luminosity spread.  Finally, the two
distributions are offset in their absolute flux scales by nearly an
order of magnitude. Most of this offset is due to different scalings
of the luminosity with spin parameters: our best-fit luminosity law
scales as $\dot E^{1/2}$, whereas Hartman et al.\ assume $\dot
E^{1/3}$. If the latter dependence were adopted for our beamed
model, the resulting flux distribution would shift to the left,
approaching the smeared pseudo-luminosity distribution.

To determine whether our beaming model alone produces a flux
distribution sufficiently wide to describe the data, we have
evaluated the likelihood for population models in which the relevant
flux depends not only on the purely geometrical spread due to
viewing orientation, but also on an additional, intrinsic spread in
luminosity for objects with a given $P$ and $\dot P$. We modeled the
intrinsic luminosity by a log-normal distribution with mean
set by the scale parameter $L_0$ and coefficients $\alpha$ and
$\beta$ as in Eq.~\ref{lr.eq}, and with width equal to $\epsilon$
times the mean, where we treated $\epsilon$ as a parameter to be 
inferred. We examined the likelihood function as a function of
$\epsilon$ and $L_0$, holding the other parameters fixed. We found
that the likelihood decreases monotonically for all values $0 \le
\epsilon \le 1$. Although we have not carried out a thorough search
of parameter space for a global maximum of the likelihood, we
interpret this behavior as evidence that geometric orientation 
effects are sufficient, without need for ``hidden'' variables, to
account for the full spread in observed fluxes manifested by the
radio pulsar population.

To provide a degree of separation between our luminosity model and
the results of the likelihood analysis, we may also simply drop flux
information from the likelihood function, by ignoring $\Db$
observables (Sec.~\ref{like.sec}) and maximizing only the ${\cal
L}_a$ function, Eq.~\ref{la.eq}.  Assumptions inherent in the
luminosity model then enter into the simulation only through a given
pulsar's detectability.  We find that the velocity and luminosity
parameter values at maximum likelihood, and the inferred birthrate,
change little (with the exception, as noted in the previous section,
of a decreased average velocity) when flux and proper motion are
eliminated from the likelihood function.


We conclude that the impact of our luminosity and beam assumptions
on birth velocity results is likely confined to the effects of any
insufficiency in the empirically-derived radio beam geometry we use,
and then only indirectly.  Further exploration of the effects of
different beam properties is warranted and is underway.

\subsection{Comparison of Velocity PDF with previous results}
\label{comp.sec}

Several attempts to constrain the birth velocities of NSs have been
described in the literature. 
(See Cordes \& Chernoff 1998 for a discussion of the analyses of
Hansen \& Phinney 1997\nocite{hp97} and Lorimer et al.\
1997\nocite{lbh97}).
As emphasized earlier, our method offers
an accounting of selection effects not previously available, 
rigorous treatment of uncertainties in, e.g., proper motion and distance, 
and analysis of the {\em joint\/} distribution of observable
pulsar characteristics. Here, we compare our result to those of two
recent studies. 

Cordes \& Chernoff (1998; hereafter CC98) examined the kinematics of 49
pulsars younger than 10 Myr with measured proper motions. Through a
multi-dimensional likelihood analysis, but purposely not accounting for
selection effects, they arrived at a birth velocity distribution for
these objects described by $\sigma_{v_1} = 175^{+19}_{-24}$ km~s$^{-1}$,
$\sigma_{v_2} = 700^{+300}_{-132} $ km~s$^{-1}$, and $w_1 = 0.86$. This
distribution, shown in Fig.~\ref{vdist.fig}, significantly
under-represents the high-velocity population implied by our analysis.
To understand this difference, we have investigated the selection
effects inherent in the choice of the 49 pulsars analyzed by CC98:
because measurement of proper motion typically requires a time baseline
of many years, all of the CC98 pulsars were discovered in some of the
earliest pulsar surveys, which were substantially less sensitive than
present-day surveys and directed along the Galactic plane. As a result,
the CC98 pulsars, in addition to being young, constitute a
volume-limited sample. We find that, if pulsars conform to our best-fit
luminosity law, the detection efficiency of
low-latitude, volume-limited surveys drops steadily with increasing
velocity, so that at 1000 km~s$^{-1}$ it is roughly one-third of its
low-velocity value. In contrast, recent deep high-latitude (e.g., Foster
et al.\ 1995; Thorsett et al.\ 1993) and all-sky (e.g., Manchester et
al.\ 1996) surveys are sensitive to much of the volume occupied by
high-velocity pulsars (away from the disk where propagation effects
limit the survey volume). Such surveys combat the
velocity-dependent selection effects to which early pulsar catalogs were
subject. 
In this regard, an indirect
observational bias against high-velocity pulsars remains unaccounted for
in our analysis:  by construction, our likelihood function assumes that
proper motion measurements are available for an unbiased, representative
subset of the pulsar catalog. In reality, such measurements have been
attempted predominantly for nearby, bright pulsars that were discovered
in early low-latitude surveys.
Future proper-motion measurements of pulsars discovered in
high-latitude surveys will mitigate this bias and directly probe the
high end of the pulsar velocity distribution.

Fryer, Burrows \& Benz (1998; hereafter FBB) simulate binary evolution
including NS kicks and compare the relative numbers of the resulting
evolutionary endstates, LMXBs, HMXBs, and isolated pulsars, with
observation.  The analysis of FBB complements our own because they
separate binary from kick velocities, but they do not account for
selection effects associated with pulsar detection.  FBB conclude that
some 30\% of NS are born with essentially zero kick, with the remaining
70\% being given large (500--600 km~s$^{-1}$) kicks in addition to any
orbital velocity retained following binary disruption.  Our inferred
birth velocity distribution is not inconsistent with this result: the
low-velocity ($\sigma_{v_1} \simeq 90$ km~s$^{-1}$) component could be
produced by binary break-up. This interpretation leads to other
difficulties, however, in that the birthrate of binary endstates would be
much larger than is observed if a significant fraction of NS (such as
our $w_1 = 40$\%) received zero kick at birth (Dewey \& Cordes
1987)\nocite{dc87}.  FBB require only that their simulation yield
birthrates of binaries {\em no smaller than\/} are observed, without
penalizing models that generate too many binaries.  The
necessity for small kicks in FBB's analysis is dictated by
their adopted 1\% minimum retention fraction for globular clusters. 
Because most cluster pulsars have different evolutionary histories
than NSs in the Galactic disk (e.g., as a result of collisional
interactions and the higher binary fraction within clusters), birth
velocity distributions relevant
to disk pulsars may not apply to clusters (see Sec.~\ref{esc.sec}).
We suggest that the zero-kick fraction derived by FBB, 30\%, is 
an upper bound, and that our 90 km~s$^{-1}$
component is not due solely to relict orbital velocities. This conclusion
is supported by the work of De Donder \& Vanbeveren (1999)\nocite{dv99}, who
model the effects of binary membership of NS progenitors on the ultimate
velocity distribution of single NSs, for a variety of binary properties. They
find that the distribution of asymmetry-induced kicks dominates the space
velocity distribution, with binary effects contributing an unimportant excess
at low velocities, for any distribution of kicks with an average velocity
$\gtrsim 150$ km~s$^{-1}$.  If a significant zero-kick component exists
(e.g., the distribution suggested by FBB), most binaries survive and 
few low-velocity single pulsars remain.

\section{Discussion}

\subsection{Asymmetric Kick Mechanisms}
\label{mech.sec}

It is widely believed that the large velocities imparted to neutron stars at
birth arise through a combination of orbital disruption and a ``kick'' reaction
to an asymmetric supernova explosion. Observational evidence for a kick impulse
is found in a variety of systems: misalignments between the spin and orbital
angular momentum axes of the relativistic NS-NS binaries B1913+16 and B1534+12
(Wex, Kalogera \& Kramer 2000; Arzoumanian, Taylor \& Wolszczan 
1998)\nocite{wkk00,atw98} and the present-day
spin-orbit configuration of the binary pulsar
J0045$-$7319 (Kaspi et al.\ 1996)\nocite{kbm+96}.  Also, direct measurement of
the proper motions of non-binary radio pulsars yield, in extreme
cases, velocities far greater than can be provided by orbital motion,
$\gtrsim 1000$ km~s$^{-1}$.  The Guitar Nebula pulsar, e.g., has a projected
(two-dimensional) velocity $\sim 1600$ km~s$^{-1}$, oriented nearly parallel 
to the Galactic plane (Cordes et al.\  1993)\nocite{crl93}. 
Finally, the survival of binary systems into late
evolutionary stages, e.g., NS-NS, NS-white dwarf, or NS-black hole binaries,
depends on the magnitude, frequency of occurrence, and preferred orientation
(if any) of asymmetric kicks. The rates of survival of such systems have been
estimated, without quantitative knowledge of the NS kick velocity distribution,
through population synthesis models including orbital evolution (e.g., Dewey \&
Cordes 1987)---such efforts have affirmed the need for a velocity impulse on
timescales shorter than or comparable to the orbital motion. 

The kick magnitudes required by observations impose stringent
constraints on theoretical models.  No single mechanism for producing
kicks as large as 1500 km~s$^{-1}$ has been conclusively identified; see
Lai, Chernoff \& Cordes (2001; hereafter LCC)\nocite{lcc01} for a recent review of
proposed mechanisms. Other clues to the nature of kicks have been
scarce---the timescale on which kicks act is not known, and no
correlation has been firmly established between the magnitudes or
preferred directions of kicks and any other pulsar property (Deshpande,
Ramachandran \& Radhakrishnan 1999)\nocite{drr99}.  A possible exception
is suggested by recent {\em Chandra\/} results for the Vela and Crab
pulsars that show near alignment of the projected proper motion vectors
and spin axes for these two young objects (but see Wex et al.\
2000\nocite{wkk00} for a possible counter-example). LCC explore the implications
of such a correlation on proposed kick mechanisms.

For prompt natal kicks, two models have been explored in some detail.
Large-scale density asymmetries in a presupernova core, seeded perhaps by
energetic convection in inner burning shells driving overstable g-mode
oscillations, can produce large kicks as a result of non-isotropic shock
propagation and breakout during the explosion (see Lai 1999\nocite{lai99} for a
review). The role of rotational averaging of the momentum impulse, however, 
is unclear---it may be difficult to produce spin-aligned kicks in this model
(LCC)---and the existence of global asymmetries has not been verified
numerically, as computational limitations have restricted simulations so
far to consideration of two-dimensional models. Asymmetric emission of
neutrinos in the presence of a strong magnetic field (due ultimately to
CP violation in the weak interaction) can also impart momentum to a
proto-NS, but impulses larger than a few hundred km~s$^{-1}$ 
require very large magnetic fields ($10^{15-16}$ Gauss; Arras \& Lai
1999)\nocite{al99}.

An alternative to large impulsive kicks at birth is the electromagnetic 
rocket effect
suggested by Harrison \& Tademaru (1975), which would accelerate a
pulsar with an off-center magnetic dipole gradually over its initial
spin-down timescale. This mechanism encounters two difficulties.
Gravitational radiation due to unstable r-mode oscillations (e.g.,
Anderson 1998) may considerably enhance spin-down of a newborn neutron
star so that the rocket effect acts on a much shorter timescale,
resulting in a lower final velocity.  Also, sufficiently large
velocities may not be achievable on timescales short compared to the
orbital periods of the progenitor systems that produce close NS-NS
binaries or systems like PSR J0045$-$7319. 

The significant fraction of the birth velocity distribution lying above
500 km~s$^{-1}$ cannot have its origin in
presupernova orbital motion, strong evidence for asymmetric kicks
through a mechanism that must be able to produce velocities at least as
high as 1500 km~s$^{-1}$.  On the other hand, our results also suggest
that up to 40\% of newborn NSs experience only a small, or
zero, kick. It seems likely, then, that two mechanisms are needed, or
that the mechanism producing large kicks (e.g., global hydrodynamic
asymmetry) acts only in a subset of objects, perhaps through a threshold
in the explosion energy or the spin rate of the progenitor core. As
discussed above, kicks are needed to avoid producing
too-many binary systems containing NSs, suggesting that some mechanism
(e.g., asymmetric neutrino emission) responsible for small kicks is
active in nearly all instances of core collapse. A bimodal birth
velocity distribution is, however, difficult to understand as a
combination of independent mechanisms that produce large and small
kicks (or a large kick and an uncorrelated residual velocity from 
disruption of an orbit). The convolution of two such independent
processes should result in a broad {\em uni-modal\/} birth velocity
distribution. Our analysis demonstrates, however, a clear preference
for a distribution that can be adequately described (relative to any
single-Gaussian distribution) as a combination of two Gaussian
components with very different dispersions.

\subsection{Escape from the Gravitational Potentials of the Galaxy and
Globular Clusters}
\label{esc.sec}

Leonard \& Tremaine (1990)\nocite{lt90} derive a lower bound of $V_e =
430$ km~s$^{-1}$ for the escape velocity from the Galactic potential in
the solar neighborhood; the value of $V_e$ is larger at smaller Galactic
radii. Assuming a typical escape velocity of 500 km~s$^{-1}$, we find
that 50\% of pulsars will be born with sufficient velocity to escape from
the Galaxy for our two-component velocity model, or nearly 40\% for the
one-component model. The larger fraction is comparable to that derived
by Lyne \& Lorimer (1994; hereafter LL94)\nocite{ll94} based on their
one-component
distribution. CC98 derived an unbound fraction of 25\%, but argued that
this figure was likely underestimated by a factor of two due to
selection against high-velocities in their pulsar sample.

Drukier (1996)\nocite{dru96} estimates the fraction of newborn NSs
retained by globular clusters as a function of kick velocity, for both
single stars and binaries containing NSs. It is generally believed
that clusters must retain at least 10\% of the NS born within them, but
$V_e$ for a globular cluster
is just $\sim 30$ km~s$^{-1}$. For the LL94 distribution
Drukier finds that at most 4\% of NSs are retained for the most massive 
clusters, with typical fractions much smaller.  
Comparison of our preferred birth velocity distribution with LL94
suggests that the latter significantly under-represents the low-velocity
population of NS
(as also suggested, e.g., by Hartman et al.\ 1997\nocite{hbwv97}): in our
preferred model, the fraction of objects
with birth velocity $\lesssim 30$ km~s$^{-1}$ is an order of magnitude
larger than that for the LL94 distribution. 
To the extent that our conclusions for pulsars born in the disk are
applicable to NSs in globular clusters, the enhanced low-velocity
population in our model relative
to LL94 should significantly increase the cluster retention rate
over Drukier's findings. If our low-velocity component includes a significant
contribution from pre-disruption orbital speeds, 
i.e., in the limit of zero kick for $w_1 = 40$\% of the population, the
retention fraction is increased further,
perhaps to as large as several tens of percent. 


\subsection{Pulsar-Supernova Remnant Associations}
\label{snr.sec}

The supernova blast that accompanies NS birth drives a fast shock wave
in the external medium. Our results for the birth velocity distribution
of NSs may be used to estimate the fraction of objects that lie
within their host supernova remnants (SNR), an important criterion 
used to assess proposed NS-SNR associations. We assume an explosion
energy of $10^{51} E_{51}$ ergs in a medium of hydrogen density
$n_0$ cm$^{-3}$.  For Sedov-Taylor expansion (Shull \& Draine
1987)\nocite{sd87}, the remnant scale at time $10^4 t_4$ yrs is $R_s =
12.5\;(E_{51}/n_0)^{1/5} t_4^{2/5}$ pc. A NS with birth velocity $v_3
10^3$ km~s$^{-1}$ passes through the edge of the remnant at $t_4 = 1.45\;
(E_{51}/n_0)^{1/3} v_3^{-5/3}$. 
The shock becomes radiative (Shull 1987)\nocite{shu87} at later times,
$t_{\rm sh} \simeq 1.91\;
E_{51}^{3/14} n_0^{-4/7} 10^4$ yr, when the size is $R_{\rm sh}  = 16.2\;
E_{51}^{2/7}n_0^{-3/7}$ pc. Thence, $R = R_{\rm sh} (t/t_{\rm
sh})^{2/7}$ until the remnant slows to about 20 km~s$^{-1}$ and merges
with the interstellar medium. In this regime, $t_4 = 1.51\;E_{51}^{11/35}
n_0^{-13/35} v_3^{-7/5}$. 

\begin{figure}
\epsscale{0.8}
\plotone{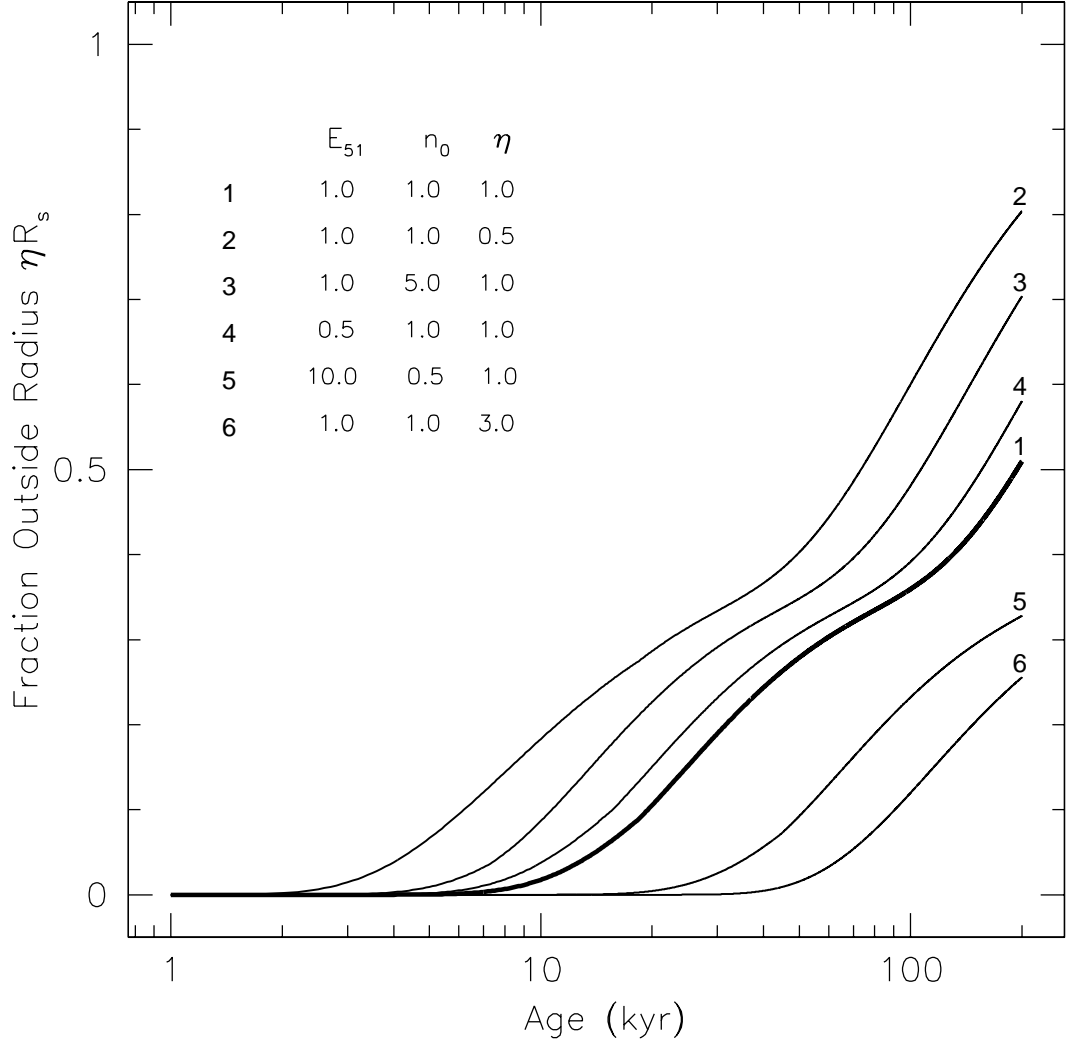}
\caption{\label{snr.fig} 
The fraction of neutron stars seen to be
outside a characteristic radius of their host remnants (see text,
Sec.~\ref{snr.sec}). 
Some objects are physically outside but are projected to be interior.
$R_s$ is the shell radius and $\eta$ is the fraction of the
shell radius considered.  
Different curves are for different explosion energies, ambient
interstellar densities, and radius fractions. 
(1) Heavy line:  Fraction outside the shell of a SNR with
    nominal energy expanding into an ISM of unit density:
    $E_{51} = 1, n_0 = 1, \eta = 1$.
(2) Fraction outside half the shell radius of a SNR with
    nominal energy expanding into an ISM of unit density:
    $E_{51} = 1, n_0 = 1, \eta = 0.5$.
(3) Fraction outside the shell of a SNR with
    nominal energy expanding into an ISM with larger density:
    $E_{51} = 1, n_0 = 5, \eta = 1$.
(4) Fraction outside the shell of a SNR with
    half nominal energy expanding into an ISM of unit density:
    $E_{51} = 0.5, n_0 = 1, \eta = 1$.
(5) Fraction outside the shell of a hypernova with
    $\times 10$ nominal energy expanding into an underdense ISM: 
    $E_{51} = 10, n_0 = 0.5, \eta = 1$.
(6) Fraction outside three times the shell radius of a SNR with
    nominal energy expanding into an ISM of unit density:
    $E_{51} = 1, n_0 = 1, \eta = 3$.
}
\end{figure}

The CDF for birth velocities less than $v$ is equivalent to the CDF for NSs
lying within the remnant at time $t$ when $v$ and $t$ are related as above for
both the Sedov-Taylor and radiative regimes. We show these CDFs, modified 
by the effects of projection for a distant observer, in Fig.~\ref{snr.fig} 
for several combinations of explosion energy and ambient density, as
a function of NS age for our preferred birth velocity distribution. We find
that up to 10\% of pulsars younger than 20 kyr will appear to lie outside
of their host remnant shells. Such a small fraction is consistent with
the set of known and proposed pulsar-SNR associations (e.g., Gaensler \& 
Johnston 1995)\nocite{gj95}. Of course, to assess the plausibility of
a particular proposed association, age, distance, and especially
proper motion measurements must be made. 

\subsection{LIGO Rates and Binary Inspiral Sources of GRBs}

Both the number and orbital characteristics of binaries that survive a
supernova explosion depend sensitively on the distribution of kick magnitudes.
A number of studies (e.g., Dewey \& Cordes 1987\nocite{dc87}; 
Wijers, van Paradijs \& van den Heuvel 1992\nocite{wvv92}; 
Brandt \& Podsiadlowski 1995\nocite{bp95}; 
Portegies Zwart \& Spreeuw 1996\nocite{ps96};
Lipunov et al.\ 1997\nocite{lpp97}; 
Bloom, Sigurdsson \& Pols 1999\nocite{bsp99}) 
have shown that the
birthrates of double neutron star (DNS) and low- and high-mass X-ray binary
systems decrease with increasing characteristic kick velocity. For the DNS
binaries, the resulting orbital period and eccentricity also determine
the rate at
which gravitational radiation is emitted; high-eccentricity and short-period
systems will typically merge within $10^9$ yr, so that the rate of
merger events will
be smaller than the birthrate of all DNS binaries. In their final moments,
merging systems are expected to produce gravitational radiation detectable by
LIGO.  

Assumed quantities (such as the importance of the high end of the initial mass
function or the supernova rates in nearby host galaxies) contribute some
uncertainty to DNS birthrate estimates, but the greatest source of uncertainty
has been the unknown kick velocity distribution, yielding rates
spanning in some cases more than an order of magnitude (e.g., from 9 to 384
Myr$^{-1}$ per galaxy in the work of Portegeis Zwart \& Spreeuw
1996\nocite{ps96}).
For the highest
kick velocities that they model (a Maxwellian distribution with 3-dimensional
velocity dispersion of 450 km~s$^{-1}$, their Model C), Portegeis Zwart \&
Yungelson (1998) find a DNS birthrate of 17 Myr$^{-1}$ and a merger rate over
10 Gyr of 12 Myr$^{-1}$. They also find that NS-black hole mergers will occur
at a rate of 1 per Myr.  Similarly, Bloom, et al.\ 
(1999)\nocite{bsp99} find a DNS
birthrate of 3 Myr$^{-1}$ for their simulations incorporating large kicks
(drawn from a Maxwellian distribution with one-dimensional dispersion of 270 km
s$^{-1}$).  The birth velocity
distribution that we have derived should improve such estimates significantly.
Belczy\'{n}ski \& Bulik (1999)\nocite{bb99} provide an expression linking the
rate of DNS mergers to the kick velocity in one or more Gaussian
components based on their binary evolution simulation, from which we
infer a merger rate of 30 Myr$^{-1}$. 
A general feature of two-component kick velocity distributions is that
the resulting merger rate will be dominated by the small-kick component,
but systems that remain bound following larger kicks also contribute
through their shorter merger times owing to their high orbital
eccentricities (see, e.g., Table 1 of BPS). The inclusion
of a significant fraction of large kicks into binary evolution
simulations will likely bring their birth and merger rate predictions in
line with empirical estimates based on the small sample of known DNS
systems (e.g., Arzoumanian, Cordes \& Wasserman 1999\nocite{acw99}; 
Kalogera et al.\ 2001\nocite{knst01}).

Models linking the coalescence of compact binaries with (at least some)
gamma-ray bursts (GRBs) have been developed
(Pacy\'{n}ski 1986\nocite{pac86}; Goodman 1986\nocite{goo86}), 
and can
account for many properties of GRBs, including their energetics, spatial
distribution, and (depending on assumptions about beaming) event rate.
In this context, the discussion of DNS merger rates above applies
equally well to GRBs---our inferred birth velocity distribution supports
published estimates of the merger rate that assume the largest kicks.
Bloom et al.\  (1999)\nocite{bsp99} model the effects of
momentum impulses
on the systemic motion of a compact binary in the gravitational
potential of its host galaxy to determine the expected range of offsets
on the sky between a GRB and the host galaxy. They find that the details
of the distribution of kicks imparted to NSs at birth do not
significantly affect these offsets because large kicks tend to disrupt
the binary. A system's survival therefore acts as a velocity filter so
that, for an ensemble of systems, our larger average kicks will not
necessarily increase the spatial offsets expected of coalescing binaries
containing NSs.

Binary systems containing a stellar-mass black hole are also potential
sources of gravitational radiation for the LIGO and LISA detectors,
and may drive certain gamma-ray bursts. The systemic velocities of
low-mass X-ray binaries containing black hole candidates (BHC-LMXBs) appear
to be smaller than those of NS-LMXBs (White \& van Paradijs 
1996)\nocite{wp96}. 
Black holes that form promptly upon the progenitor
star's collapse are not expected to undergo kicks, but some black holes
may form after passing through a short lived NS (or proto-NS)
phase, e.g., if accretion of fallback material 
pushes the compact object's mass over the 
Chandrasekhar limit (see, e.g., Brandt, Podsiadlowski \& Sigurdsson
1995; hereafter BPS). Such ``delayed formation'' black holes may receive
asymmetric kicks just as neutron stars do, with smaller kinematic
consequences because of the larger mass of black holes. 
One BHC-LMXB, X-ray Nova Sco 1994, does have
an unusually high peculiar velocity, $\simeq 110$ km~s$^{-1}$, as do a
few high-mass X-ray binaries. As pointed out by BPS,
formation scenarios of Nova Sco are simplified if one invokes an
asymmetric kick at birth; moreover, elemental abundances in its
companion support a supernova origin for the black hole (Israelian et
al.\ 1999)\nocite{irb+99}.  Extension of our radio-pulsar birth
velocity distribution to black-hole systems would of course be
speculative, but the properties of Nova Sco 
suggest that kicks similar
to those associated with NSs also occur as part of some (perhaps rare)
black-hole formation mechanisms.  Any such kick would have implications
for the survival rate of binaries containing a delayed-formation black
hole. It should be noted that black-hole velocity distributions inferred 
from objects in binary systems may be prone to selection bias against
high-velocity BHCs, the formation of which may have disrupted the host
binary.

\subsection{Populations of Isolated Pulsars}
\label{isol.sec}

Early expectations that large numbers of old NSs accreting from the
interstellar medium would be detectable in the soft X-ray band have not
been borne out by observations (Neuh\"{a}user \& Tr\"{u}mper
1999)\nocite{nt99}. The earliest estimates, however, were based on an
assumed NS velocity distribution peaking at $\sim 40$ km~s$^{-1}$ 
(Treves \& Colpi 1991)\nocite{tc91}. The much
larger average velocities inferred by LL94 and CC98 suggested a solution
to this discrepancy, as the accretion rate is expected to scale with
velocity as $v^{-3}$; magnetic field decay on timescales less than
$10^9$ years could also explain why so few ISM-accreting NSs are seen
(for a review, see Treves et al.\ 2000)\nocite{ttzc00}. As described
above, however, incomplete accounting for selection effects (those
arising from radio-wave propagation in the Galactic disk) likely leads
to an underestimated low-velocity proportion in the analyses of LL94,
CC98, and others, potentially undermining the conclusion that the deficit
of old accreting NSs is due to higher-than-expected characteristic
velocities. Our analysis models the effects of signal propagation
and survey instrumentation, providing a more robust census of the
low-velocity population. We infer a larger low-velocity
fraction than do LL94 and CC98, a result which, together with our limit
on an additional very low velocity component ($< 5$\% with birth
velocity $< 50$ km~s$^{-1}$), should firmly constrain theoretical
expectations for the number of nearby isolated NSs accreting from the ISM. 
As pointed out by Neuh\"{a}user \& Tr\"{u}mper (1999)\nocite{nt99}, the
number of isolated NS detections in the ROSAT All-Sky Survey is
consistent with estimates made by Madau \& Blaes (1994)\nocite{mb94}
based on a low birth-velocity distribution modified by
diffusive heating of the NSs in the Galaxy---the resulting 
distribution peaks near 90 km~s$^{-1}$, similar to the
low-velocity component of our preferred birth velocity model.

The observational phenomena known as soft-gamma repeaters (SGRs) and
anomalous X-ray pulsars (AXPs) are believed to be associated with 
neutron stars. SGRs and AXPs are similar in many ways: high ($\sim 
10^{14}$ G ``magnetar'') canonical field strengths, slow rotation (periods 
in the range 6--12 s), and luminosities that are too large to be due to 
magnetic dipole braking alone; alternate sources of energy are field decay 
and accretion of either fallback material from the supernova explosion or 
ejecta in the vicinity of the NS (e.g., Chatterjee, Hernquist \& Narayan 
2000; van Paradijs, Taam \& van den Heuvel 1995)\nocite{chn00,pth95}.
In addition to their distinct high-energy emission characteristics,
a possibly significant observational difference between the two types of
objects is that most AXPs are found near the centers of supernova
remnants (so that associations are quite certain), while SGRs are found
far from SNR centers, making associations less likely and requiring
large ($\sim 1000$ km~s$^{-1}$) NS velocities if the associations are to
be viable (Gaensler et al.\ 2001)\nocite{gsgv01}.
Our results suggest that such high
velocities are not rare among radio pulsars, certainly, and possibly
among all NSs. 
Although distinctions between AXPs and SGRs on purely kinematic grounds
remain controversial, models that invoke episodic interaction (e.g.,
accretion or spin-down due to the propeller effect) with fall-back disks
or clumpy ejecta will clearly be influenced by the assumed velocities of
the young neutron stars.

SGRs and AXPs may represent as much as half of the Galactic NS population 
(Gotthelf \& Vasisht
1999)\nocite{gv99}, yet they are distinct from radio pulsars in their spin
characteristics and energetics, with no apparent overlap between the two
populations.
Estimates of the NS birthrate based on radio pulsars, including our own,
must therefore be revised upward to account for these objects.

Our derived birthrate applies to isolated, rotation-powered, radio
emitting pulsars, although the radio beam need not be visible to
Earth-bound observers. Requiring non-binarity for our data sample
results in the exclusion of just four pulsars that otherwise meet
all of our selection criteria, suggesting that our birthrate
essentially accounts also for radio pulsars in long-lived,
non-interacting binary systems (selection effects that reduce survey
sensitivities for binaries are unimportant for
long-period pulsars). From analyses of the local pulsar population,
Lyne et al.\ (1998)\nocite{l+98} derive a Galactic pulsar birthrate of one
every 60 to 330 yrs, while Hartman et al.\ (1997)\nocite{hbwv97} find
model-dependent rates that cluster around (300 yr)$^{-1}$.  Our
birthrate estimate is a factor 2.5 lower still.
Much of the discrepancy between our value and that
of Hartman et al.\ (1997), however, can be attributed to the different
effective beaming fractions in our analyses, evident in
Fig.~\ref{pflux.fig} (Sec.~\ref{assump.sec}), suggesting perhaps
that the low end
of the Lyne et al.\ (1998) range is more appropriate. 
As emphasized earlier, our inferred birthrate depends 
on the assumed beam model and spin-down law (with no field decay),
and cutoffs in the distribution of magnetic field strengths at birth.
Disagreement between birthrate determinations is also likely due to
different assumed radial distributions of birthsites in the Galaxy
(Sec.~\ref{birth.sec}).  It is noteworthy that the majority of young
cataloged radio pulsars 
were discovered in high-frequency ($\nu \sim 1.4$ GHz) surveys of
the Galactic Plane, and typically not detected, because of
propagation effects, by the 0.4 GHz surveys we have simulated. A
future extension of our simulation to include high-frequency surveys
will more directly probe the population of young pulsars near their
birthsites, providing valuable constraints on their spatial
distribution.

A number
of radio-quiet but apparently rotation-driven (i.e., distinct from
SGRs and AXPs) young neutron stars have recently been
discovered at X- and $\gamma$-ray wavelengths, often associated with
supernova remnants (e.g., see Brazier \& Johnston 1999\nocite{bj99} for a
compilation); these are most naturally described as pulsars whose radio
beams do not intercept our line of sight, and so such objects would also
be included in our birthrate estimate. However, Brazier \& Johnston 
(1999)\nocite{bj99}
derive a NS birthrate $\sim (91$ yr$)^{-1}$ by counting such objects
and known radio pulsars within a distance of 3.5 kpc and less than 20
kyr old, and assuming a Gaussian radial distribution in the Galaxy.
Again, our radio pulsar birthrate is substantially lower, raising
the possibility that some young neutron stars may be truly radio quiet,
i.e., not emitting radio beams. Such a conclusion, which cannot be
ruled out at present, would be strengthened if a significant discrepancy
between the birthrates of radio pulsars and young, radio-quiet NSs
persists in future analyses of the Galactic neutron star population.

\section{Summary}

Together with adjustable models of the birth, evolution, and beaming
characteristics of the Galactic radio pulsar population, we have
simulated eight large-area surveys, and accounted for observational
selection effects, to infer the birth properties
(velocity, period, magnetic field strength) of pulsars and their
present-day radio luminosities. 
Our results suggest that the scaling of radio luminosity with
spin parameters $P$ and $\dot P$ is similar to the scaling of 
voltage drop across the magnetic polar cap in pulsar magnetosphere
models, and that old pulsars convert most of their spin-down energy
into radio emissions. For birth velocities, we find that two-component
distributions are preferred over one-component distributions, and we
infer maximum likelihood velocities for the two (three-dimensional)
Gaussian components of 90 km~s$^{-1}$ and 500 km~s$^{-1}$, with the
population
roughly equally divided between the two components. This result strongly
supports existing evidence that neutron stars are subject to ``kick''
impulses at birth, presumably through asymmetric supernova explosions,
and that the kick mechanism must be able to produce velocities of at
least 1000 km~s$^{-1}$. We find that half of all pulsars born near or
outside the solar circle will likely escape the Galaxy's gravitational
pull. The large kicks implied by our birth velocity distribution also
have important consequences for the survival rates, and final orbital
configurations, of binaries in which a neutron star is formed, as well
as for the plausibility of proposed pulsar-supernova remnant associations.

The catalog data that we have used to constrain our models provide spin
parameters, fluxes, and Galactic locations for the known pulsars, but
little in the way of direct velocity measurements through proper motions
and accurate distances. 
Very Long Baseline Interferometry techniques now being perfected (e.g.,
Fomalont et al.\ 1999\nocite{fgbc99}, Chatterjee et al.\ 2001\nocite{ccl+01}) 
will provide high-quality
proper-motion and parallax measurements for pulsars within $\sim 1$ kpc,
a substantial improvement to the current dataset. Future statistical
analyses similar to the one presented here therefore promise important gains
in our understanding of the velocity distribution of
neutron stars.

\acknowledgements
We thank J. Arons, R. Bandiera, D. Lai, T. Loredo, D. R. Lorimer, A. G.
Lyne, E. S.
Phinney, M. Ruderman, and I. Wasserman for many fruitful conversations.
C. Dolan and M. Jarvis contributed to the early development of
simulation software.  This research was conducted using the resources of
the Cornell Theory Center, which receives funding from Cornell
University, New York State, federal agencies, and corporate partners.
Part of this work was performed while ZA held a National Research
Council Research Associateship Award at NASA-GSFC. Support from NASA and
NSF grants NAG 5-2851, NAG 5-8356, AST 92-18075, AST 95-30397, and AST
98-19931, is gratefully acknowledged.

\begin{table}
\caption{Summary of model parameters.}
\begin{center}
\begin{tabular}{lcccccc}
\tableline
\tableline
Parameter &
\multicolumn{3}{c}{Peak posterior probability value\tablenotemark{a}} &
Credible &
\multicolumn{2}{c}{Range searched} \\
&
\multicolumn{2}{c}{$n = 3$} &
$n = 4.5$ &
range &
Min. &
Max. 
\\
\tableline
$\alpha$ & 
	\multicolumn{2}{c}{\phn$-1.3$\phn} & 
	\phn$-1.6$\phn & 
	$\pm\phn0.3\phn$ &
	\phn$-2$\phd\phn\phn & 
	\phs\phn0.5\phn \\
$\beta$ & 
	\multicolumn{2}{c}{\phn\phs0.4\phn} & 
	\phn\phs0.5\phn & 
	$\pm\phn0.1\phn$ &
	\phn\phs0\phd\phn\phn & 
	\phn\phs1\phd\phn\phn \\
$\log L_0$ & 
	\multicolumn{2}{c}{\phs29.3\phn} & 
	\phs29.1\phn & 
	$\pm\phn0.1\phn$ &
	\phs27.5\phn & 
	\phs30.5\phn \\
$\overline{\rm DL}$ & 
	\multicolumn{2}{c}{\phs\phn$0.5$\phn} & 
	\phs\phn0.8\phn & 
	$\pm\phn0.3\phn$ &
	\phn$-1.2$\phn & 
	\phs\phn1.2\phn \\
$\sigma_{\rm DL}$ & 
	\multicolumn{2}{c}{\phs\phn$1.4$\phn} & 
	\phs\phn1.4\phn & 
	$\pm\phn0.2\phn$ &
	\phs\phn0\phd\phn\phn & 
	\phs\phn3.2\phn \\
$\langle \log P_0 \left[{\rm s}\right]\rangle$ &
	\multicolumn{2}{c}{\phn$-2.3$\phn} & 
	\phn$-1.9$\phn & 
	$\pm\phn0.2\phn$ &
	\phn$-2.6$\phn & 
	\phn$-1.6$\phn \\
$\sigma_{\log P_0}$ & 
	\multicolumn{2}{c}{$> 0.2$} & 
	\phs\phn0.6\phn & 
	$\pm\phn 0.2\phn$ &
	\phs\phn0.1\phn & 
	\phs\phn0.7\phn \\
$\log P_0^{\rm lo}$ (Note b) & 
	\multicolumn{2}{c}{\phn$-$4\phd\phn\phn} & 
	\phn$-$4\phd\phn\phn &
	-- &
	-- &
	-- \\
$\log P_0^{\rm hi}$ & 
	\multicolumn{2}{c}{\phs\phn0\phd\phn\phn} & 
	\phs\phn0\phd\phn\phn &
	-- &
	-- &
	-- \\
$\langle \log B_0 \left[{\rm G}\right]\rangle$ &
	\multicolumn{2}{c}{\phs$12.35$} & 
	\phs$12.35$ & 
	$\pm\phn 0.10$ &
	\phs12.00 & 
	\phs12.50 \\
$\sigma_{\log B_0}$ & 
	\multicolumn{2}{c}{\phs\phn$0.40$} & 
	\phs\phn$0.65$ & 
	$\pm\phn0.05$ &
	\phs\phn0.25 & 
	\phs\phn0.50\\
$\log B_0^{\rm lo}$ & 
	\multicolumn{2}{c}{\phs11.2\phn} & 
	\phs11.2\phn &
	-- &
	-- &
	-- \\
$\log B_0^{\rm hi}$ & 
	\multicolumn{2}{c}{\phs13.8\phn} & 
	\phs13.8\phn &
	-- &
	-- &
	-- \\
$z_0$ (pc) & 
	\multicolumn{2}{c}{160\phn\phn} & 
	220\phn\phn & 
	$\pm 40\phd\phn\phn$ &
	\phs80\phd\phn\phn & 
	250\phd\phn\phn\\
\tableline
Velocity components\tablenotemark{c} & One & Two & Two & & & \\
\tableline
\rule{0em}{2.5ex}$\sigma_{v_1}$ (km~s$^{-1}$)& $290\pm30$ & \phn$90^{+20}_{-15}$ &
	$120^{+20}_{-15}$ & & \phs50\phd\phn\phn & 400\phd\phn\phn \\
\rule{0em}{2.5ex}$\sigma_{v_2}$ (km~s$^{-1}$)& --  & $500^{+250}_{-150}$ &
	$650^{+250}_{-150}$ & & 400\phn\phn & 1400\phd\phn\phn\phn \\
\rule{0em}{2.5ex}$w_1$ 	& --  & $0.4\pm0.2$ & $0.45\pm0.2$ & &
	\phs\phn0\phd\phn\phn & \phn\phn1\phd\phn\phn \\
\tableline
$\ln(\Lpsr/\Lpsr_{\rm max})$ & $-25.6$ & 0 & $-8.1$ & & & \\
\tableline
\end{tabular}
\tablenotetext{a}{The parameter value for the maximum likelihood
(eq.~\ref{like.eq}) model, and with flat ``priors'' assumed.}
\tablenotetext{b}{Lower and upper cutoffs of the log-normal distributions
for $P_0$ and $B_0$ were fixed at the values shown.}
\tablenotetext{c}{All trial distributions were truncated above 3000 km
s$^{-1}$.}
\end{center}
\end{table}

\end{document}